\documentclass{article}
\usepackage{spconf,amsmath,graphicx}

\usepackage[hidelinks]{hyperref}
\usepackage[sort]{cite}
\usepackage{pbalance}

\usepackage{acronym}
\usepackage{siunitx}
\usepackage{multirow}
\usepackage{xcolor}
\usepackage{subcaption}
\definecolor{scoreRed}{RGB}{220,50,47}
\definecolor{scoreYellow}{RGB}{255,235,59}
\definecolor{scoreGreen}{RGB}{0,150,0}
\definecolor{corrRed}{RGB}{178,24,43}
\definecolor{corrWhite}{RGB}{247,247,247}
\definecolor{corrBlue}{RGB}{5,48,97}
\definecolor{disRed}{RGB}{139,0,0}
\definecolor{disYellow}{RGB}{255,204,0}
\definecolor{disGreen}{RGB}{51,170,0}
\definecolor{refGreen}{RGB}{0,150,0}
\definecolor{refYellow}{RGB}{255,235,59}
\definecolor{refRed}{RGB}{220,50,47}

\newcommand{\getscorecolor}[1]{%
    \ifdim#1pt<1.2pt scoreRed%
    \else\ifdim#1pt<1.4pt scoreRed!95!scoreYellow%
    \else\ifdim#1pt<1.6pt scoreRed!90!scoreYellow%
    \else\ifdim#1pt<1.8pt scoreRed!85!scoreYellow%
    \else\ifdim#1pt<2.0pt scoreRed!80!scoreYellow%
    \else\ifdim#1pt<2.2pt scoreRed!70!scoreYellow%
    \else\ifdim#1pt<2.4pt scoreRed!60!scoreYellow%
    \else\ifdim#1pt<2.6pt scoreRed!50!scoreYellow%
    \else\ifdim#1pt<2.8pt scoreRed!40!scoreYellow%
    \else\ifdim#1pt<3.0pt scoreRed!30!scoreYellow%
    \else\ifdim#1pt<3.2pt scoreYellow%
    \else\ifdim#1pt<3.4pt scoreYellow!90!scoreGreen%
    \else\ifdim#1pt<3.6pt scoreYellow!80!scoreGreen%
    \else\ifdim#1pt<3.8pt scoreYellow!70!scoreGreen%
    \else\ifdim#1pt<4.0pt scoreYellow!60!scoreGreen%
    \else\ifdim#1pt<4.2pt scoreYellow!50!scoreGreen%
    \else\ifdim#1pt<4.4pt scoreYellow!40!scoreGreen%
    \else\ifdim#1pt<4.6pt scoreYellow!30!scoreGreen%
    \else\ifdim#1pt<4.8pt scoreYellow!20!scoreGreen%
    \else\ifdim#1pt<5.0pt scoreYellow!10!scoreGreen%
    \else scoreGreen%
    \fi\fi\fi\fi\fi\fi\fi\fi\fi\fi\fi\fi\fi\fi\fi\fi\fi\fi\fi\fi%
}

\newcommand{\getcorrcolor}[1]{%
    \ifdim#1pt<-0.9pt corrRed%
    \else\ifdim#1pt<-0.8pt corrRed!95!corrWhite%
    \else\ifdim#1pt<-0.7pt corrRed!90!corrWhite%
    \else\ifdim#1pt<-0.6pt corrRed!80!corrWhite%
    \else\ifdim#1pt<-0.5pt corrRed!70!corrWhite%
    \else\ifdim#1pt<-0.4pt corrRed!60!corrWhite%
    \else\ifdim#1pt<-0.3pt corrRed!50!corrWhite%
    \else\ifdim#1pt<-0.2pt corrRed!40!corrWhite%
    \else\ifdim#1pt<-0.1pt corrRed!30!corrWhite%
    \else\ifdim#1pt<0pt corrRed!20!corrWhite%
    \else\ifdim#1pt<0.1pt corrWhite%
    \else\ifdim#1pt<0.2pt corrWhite!80!corrBlue%
    \else\ifdim#1pt<0.3pt corrWhite!70!corrBlue%
    \else\ifdim#1pt<0.4pt corrWhite!60!corrBlue%
    \else\ifdim#1pt<0.5pt corrWhite!50!corrBlue%
    \else\ifdim#1pt<0.6pt corrWhite!40!corrBlue%
    \else\ifdim#1pt<0.7pt corrWhite!30!corrBlue%
    \else\ifdim#1pt<0.8pt corrWhite!20!corrBlue%
    \else\ifdim#1pt<0.9pt corrWhite!10!corrBlue%
    \else\ifdim#1pt<1.0pt corrBlue!90!corrWhite%
    \else corrBlue%
    \fi\fi\fi\fi\fi\fi\fi\fi\fi\fi\fi\fi\fi\fi\fi\fi\fi\fi\fi\fi%
}

\newcommand{\getdiscolor}[1]{%
    \ifdim#1pt<-4.75pt disGreen%
    \else\ifdim#1pt<-4.5pt disGreen!90!disYellow%
    \else\ifdim#1pt<-4.25pt disGreen!80!disYellow%
    \else\ifdim#1pt<-4.0pt disGreen!70!disYellow%
    \else\ifdim#1pt<-3.75pt disGreen!60!disYellow%
    \else\ifdim#1pt<-3.5pt disGreen!50!disYellow%
    \else\ifdim#1pt<-3.25pt disGreen!40!disYellow%
    \else\ifdim#1pt<-3.0pt disGreen!30!disYellow%
    \else\ifdim#1pt<-2.75pt disGreen!20!disYellow%
    \else\ifdim#1pt<-2.5pt disGreen!10!disYellow%
    \else\ifdim#1pt<-2.25pt disYellow!90!disGreen%
    \else\ifdim#1pt<-2.0pt disYellow!80!disRed%
    \else disRed%
    \fi\fi\fi\fi\fi\fi\fi\fi\fi\fi\fi\fi%
}
\newcommand{\getrefcolor}[1]{%
    \ifdim#1pt<0.15pt refGreen
    \else\ifdim#1pt<0.25pt refGreen!78!refYellow
    \else\ifdim#1pt<0.35pt refGreen!56!refYellow
    \else\ifdim#1pt<0.45pt refGreen!33!refYellow
    \else\ifdim#1pt<0.55pt refGreen!11!refYellow
    \else\ifdim#1pt<0.65pt refYellow!89!refRed
    \else\ifdim#1pt<0.75pt refYellow!67!refRed
    \else\ifdim#1pt<0.85pt refYellow!44!refRed
    \else\ifdim#1pt<0.95pt refYellow!22!refRed
    \else\ifdim#1pt<1.0pt refRed
    \else refRed
    \fi\fi\fi\fi\fi\fi\fi\fi\fi\fi%
}

\newcommand{\scorebar}[1]{#1\hspace{1pt}\textcolor{\getscorecolor{#1}}{\rule{3pt}{1ex}}}
\newcommand{\corrbar}[1]{#1\hspace{1pt}\textcolor{\getcorrcolor{#1}}{\rule{3pt}{1ex}}}

\newcommand{\refbar}[1]{#1\hspace{1pt}\textcolor{\getrefcolor{#1}}{\rule{3pt}{1ex}}}

\usepackage{cite}
\usepackage{amsmath,amssymb,amsfonts}
\usepackage{algorithmic}
\usepackage{textcomp}
\usepackage{booktabs} 

\title{ACOUSTIC TELEPORTATION VIA DISENTANGLED NEURAL AUDIO CODEC REPRESENTATIONS}
\twoauthors
 {Philipp Grundhuber, Mhd Modar Halimeh}
	{Fraunhofer Institute for Integrated Circuits (IIS)\\ Erlangen, Germany}
 {Emanuël Habets}
	{International Audio Laboratories Erlangen\\ Erlangen, Germany}

\name{Philipp Grundhuber$^{\dagger}$ \qquad Mhd Modar Halimeh$^{\dagger, \S}$ \thanks{\textsuperscript{$\S$}Mhd Modar Halimeh is now with Starkey Hearing Technologies, Eden Prairie, MN, US}
\qquad Emanu{\"e}l A. P. Habets$^{\star}$}
\address{$^{\dagger}$ Fraunhofer Institute for Integrated Circuits (IIS), Erlangen, Germany \\ $^{\star}$ International Audio Laboratories Erlangen\textsuperscript{$\ast$}, Erlangen, Germany \thanks{\textsuperscript{$\ast$}A joint institution of Fraunhofer IIS and Friedrich-Alexander-Universit{\"a}t Erlangen-N{\"u}rnberg (FAU), Germany.}} 

\acrodef{AT}[AT]{Acoustic Teleportation}
\acrodef{RIR}[RIR]{room impulse response}
\acrodef{RVQ}[RVQ]{residual vector quantization}
\acrodef{NAC}[NAC]{Neural Audio Codec}
\acrodef{PCA}[PCA]{principal component analysis}
\acrodef{ZS-TTS}[ZS-TTS]{zero-shot text-to-speech synthesis}

\begin{document}
\ninept
\maketitle
\begin{abstract}
This paper presents an approach for acoustic teleportation by disentangling speech content from acoustic environment characteristics in neural audio codec representations. \textit{Acoustic teleportation} transfers room characteristics between speech recordings while preserving content and speaker identity. We build upon previous work using the EnCodec architecture, achieving substantial objective quality improvements with non-intrusive ScoreQ scores of 3.03, compared to 2.44 for prior methods. Our training strategy incorporates five tasks: clean reconstruction, reverberated reconstruction, dereverberation, and two variants of acoustic teleportation. We demonstrate that temporal downsampling of the acoustic embedding significantly degrades performance, with even 2× downsampling resulting in a statistically significant reduction in quality. The learned acoustic embeddings exhibit strong correlations with RT60. Effective disentanglement is demonstrated using t-SNE clustering analysis, where acoustic embeddings cluster by room while speech embeddings cluster by speaker.
\end{abstract}

\begin{keywords}
neural audio coding, disentanglement learning
\end{keywords}
\section{Introduction}
Audio codecs compress audio signals into discrete codes for efficient transmission and subsequent reconstruction \cite{Spanias19941541}. Conventional codecs rely on engineered signal processing blocks \cite{valin2012rfc, neuendorf2013iso}, while \acp{NAC} have emerged as powerful alternatives showing high decoded audio quality with increased bitrate efficiency \cite{AudioDEC, defossez2022highfi, DAC}. The compressed representations learned by these neural approaches have proven valuable beyond simple compression, enabling downstream applications like \ac{ZS-TTS} \cite{wang2023neural, zhang2024speechtokenizer}. 

Recent works have explored disentangled latent spaces in \acp{NAC}. These methods partition the latent embedding space to isolate specific types of information. SD-Codec \cite{bie2025learning} assigns different sources (speech, music, and sound effects) to distinct \ac{RVQ} codebooks. The explicit disentanglement of sources across specific quantizers enables reconstruction of both individual sources and their mixtures. SRCodec \cite{zheng2024srcodec} uses split-\ac{RVQ} to separate lower- and higher-dimensional speech features. Speech is often decomposed into three physical components: timbre, prosody, and content information, enabling efficient speech processing with lower token usage while maintaining performance in reconstruction and voice conversion tasks. Methods include autoencoders \cite{chan2022speechsplit2,SpeechTripleNet}, \ac{NAC} approaches like FreeCodec \cite{zheng2024freecodec}, and factorized vector quantization as used in NaturalSpeech 3 \cite{NaturalSpeech3}. Omran et al. \cite{omran2023disentangling} present an approach for separating speech signals from environmental characteristics in partitioned embedding spaces of SoundStream \cite{zeghidour2021soundstream}. One partition represents speech content, while others either capture acoustic information or additive noise. For noise separation, they allocate equal embedding dimensions to speech and noise components. For reverberation disentanglement, the acoustic embedding is temporally downsampled by a factor of 10. While their approach demonstrates reasonable disentanglement of additive background noise and reverberation from speech, the quality of the manipulated signals, e.g., dereverberated signals, is limited and the output suffers from audible artifacts. 

Our primary contribution is to investigate "\ac{AT}": extracting acoustic information in the form of an acoustic embedding from one recording and applying it to speech recorded in a different environment. This transfers speakers between acoustic spaces while preserving speech content. We extend Omran et al.'s approach \cite{omran2023disentangling} using the EnCodec architecture \cite{defossez2022highfi}, which significantly increases objective performance. Additionally, we conduct an ablation study of the training strategies, incorporating five tasks that enhance model versatility and disentanglement quality. In addition, we investigate the downsampling of acoustic embeddings and its impact on the \ac{NAC} 's generalization and disentanglement capabilities. Finally, we demonstrate speaker- and room-independence of the acoustic embeddings and their correlation with the physical room parameters RT60. 

Our results show that disentangled neural codec representations effectively extract acoustic information, enabling acoustic manipulation and room characteristic estimation with applications in telecommunications, virtual acoustic environments, and speech enhancement. We provide samples of all tasks on our demo page\footnote{\url{https://www.audiolabs-erlangen.de/resources/2026-ICASSP-Acoustic-Teleportation}}.

\section{Problem Formulation}
In this work, a reverberant speech signal $x_{c,r}$ is modeled as
\begin{equation}
x_{c,r} = s_c * h_r ,
\end{equation}
where $*$ denotes the convolution operator, $s_c$ denotes an anechoic source speech signal with speech content $c$, and $h_r$ denotes a \ac{RIR} of room $r$.
\acp{NAC} typically encode audio signals into discrete embeddings that represent the input signal as entangled information from both the source speech content $s$ and the acoustic characteristics $h$, making it difficult to manipulate these aspects independently. Our objective is to learn disentangled representations where speech content and acoustic information are encoded separately. Given a recorded signal $x_{c,r}$ with anechoic speech content $c$ in room $r$, we train an encoder that produces latent representations of speech $\mathtt{s}$ and acoustics $\mathtt{h}$ such that
\begin{equation}
\{\mathtt{s}_{c,r}, \mathtt{h}_{c,r}\} = \text{Enc}(x_{c,r}) ,
\end{equation}
where the embeddings $\mathtt{s}_{c,r}$ capture ideally only speech content (such that $\mathtt{s}_{c,r} = \mathtt{s}_{c,0}$ for any room $r$) and the embeddings $\mathtt{h}_{c,r}$ capture ideally only acoustic characteristics (such that $\mathtt{h}_{c,r} = \mathtt{h}_{0,r}$ for any speech content $c$). A decoder is trained to reconstruct the original signal using
\begin{equation}
\hat{x}_{c,r} = \text{Dec}(\mathtt{s}_{c,r}, \mathtt{h}_{c,r}) .
\end{equation}

This disentanglement enables, e.g., dereverberation by setting $\mathtt{h} = 0$.

Alternatively, we can, for example, transfer the acoustic information from $x_{c_2,r_2}$ to $x_{c_1,r_1}$ where $r_1 \neq r_2$ using
\begin{align}
\{\mathtt{s}_{c_1,r_1}, \mathtt{h_{c_1,r_1}}\} &= \text{Enc}(x_{c_1,r_1}) \\
\{\mathtt{s}_{c_2,r_2}, \mathtt{h_{c_2,r_2}}\} &= \text{Enc}(x_{c_2,r_2}) . 
\end{align}
An estimate of the target signal ${x}_{c_1,r_2}$, with the anechoic speech of $x_{c_1,r_1}$ and acoustic characteristics of $x_{c_2,r_2}$, can then be obtained using
\begin{equation}
    \hat{x}_{c_1,r_2} = \text{Dec}(\mathtt{s}_{c_1,r_1}, \mathtt{h}_{c_2 ,r_2}) .
\end{equation}

\section{Proposed Method}
\subsection{Acoustic Teleportation Model}
In this paper, we use an EnCodec-based \cite{defossez2022highfi} \ac{NAC}, which operates at a sampling rate of \SI{16}{kHz}, encoder hop length of 320, a codebook size of 1024, and an output dimension of 128, where 64 coefficients are used for speech and 64 are used for the acoustic embedding. These two feature maps, each with 64 features, are then quantized separately by two \acp{RVQ} with independent codebooks, having a variable but equal number of quantizers. This renders two sets of tokens per frame: one for speech components and another for the acoustic environment information. In the original approach \cite{omran2023disentangling}, acoustic tokens are temporally downsampled by a factor of 10, i.e., effectively reducing the bitrate used for the acoustic embeddings by a factor of 10. In contrast, unless stated otherwise, we do not constrain the acoustic embeddings and assign equal bitrates to both information streams.

\subsection{Training Tasks}
Several variants of the acoustic teleportation model are trained with progressively increasing task complexity to establish the performance limits achievable by the proposed architecture. The training tasks are organized into four categories: Clean Reconstruction (CR), Reverb Reconstruction (RR), Dereverberation (DR), and Acoustic Teleportation (AT-SS, AT-DS). For clean and reverb reconstruction, no changes are made in the embedding space. Dereverberation is achieved by setting the acoustic tokens to zero. Finally, for Acoustic Teleportation, acoustic embeddings are swapped between the encoded latent spaces. This is done for latent representations from the same source (AT-SS) and from different sources (AT-DS). The tasks, including input-output mappings and embedding configurations, are summarized in Table~\ref{tab:tasks}. For an ideal encoder-decoder network with perfect disentanglement, there are 18 different pairs of speech and acoustic embeddings that are decoded into six different signals, e.g., $\mathtt{h}_{1,2}$ and $\mathtt{h}_{2,2}$ should be equal. These six output signals constitute the target signals during training.

\begin{table}[t]
\centering
\footnotesize
    \setlength{\tabcolsep}{3pt}
\begin{tabular}{llllll}
\toprule
Task & ID & Input & Speech & Acoustic & Target \\
\midrule
\multirow{1}{*}{\parbox{2cm}{Clean Reconstr.}}
& 
\multirow{1}{*}{CR} 
& $x_{c,0}$ & $\mathtt{s}_{c,0}$ & $\mathtt{h}_{c,0}$ & $x_{c,0}$ \\
\multirow{1}{*}{\parbox{2cm}{Reverb Reconstr.}} 
& \multirow{1}{*}{DR} 
 & $x_{c,r}$ & $\mathtt{s}_{c,r}$ & $\mathtt{h}_{c,r}$ & $x_{c,r}$ \\
\multirow{1}{*}{\parbox{2cm}{Dereverberation}} 
& \multirow{1}{*}{RR} 
& $x_{c,r}$ & $\mathtt{s}_{c,r}$ & $\mathbf{0}$ & $x_{c,0}$ \\
\multirow{1}{*}{\parbox{2cm}{AT Same Source}} 
& \multirow{1}{*}{AT-SS} 
& $x_{c,r_1}$ & $\mathtt{s}_{c,r_1}$ & $\mathtt{h}_{c,r_2}$ & $x_{c,r_2}$ \\
\multirow{1}{*}{\parbox{2cm}{AT Diff. Source}} 
& \multirow{1}{*}{AT-DS} 
& $x_{c_1,r_1}$ & $\mathtt{s}_{c_1,r_1}$ & $\mathtt{h}_{c_2,r_2}$ & $x_{c_1,r_2}$ \\
\bottomrule
\vspace{0cm}
\end{tabular}
\caption{Audio processing tasks, their different embedding configurations, and the corresponding target signals. Here, $c_1 \neq c_2$ and $r_1 \neq r_2$.}
\label{tab:tasks}
\end{table}

\begin{table*}[!ht]
    \caption{Performance comparison across model conditions and correlation with RT60. The number of quantizers is given by N (where "-" indicates no quantization). ScoreQ NR and ViSQOL use a color scale from 1 (red, poor) to 5 (green, excellent), while ScoreQ REF uses an inverted scale from 0.1 (green, better) to 1.0 (red, worse). Pearson correlation coefficients are color-coded from -1.0 (red, negative correlation) to 1.0 (blue, positive correlation).}
    \label{tab:performance_comparison}
    \centering
    \renewcommand{\arraystretch}{0.85}
    \setlength{\tabcolsep}{1.5pt}       
    \resizebox{\textwidth}{!}{%
    \begin{tabular}{l *{5}{c} cc ccc ccc ccc ccc r}
    \toprule
    \textbf{Model} & \multicolumn{5}{c}{\textbf{Tasks}} & \textbf{N} & \textbf{Bitrate} & \multicolumn{3}{c}{\textbf{Clean}} & \multicolumn{3}{c}{\textbf{Reverberated}} & \multicolumn{3}{c}{\textbf{Dereverberation}} & \multicolumn{3}{c}{\textbf{AT}} & {\textbf{Correlation}} \\
    \cmidrule(lr){2-6}\cmidrule(lr){9-11}\cmidrule(lr){12-14}\cmidrule(lr){15-17}\cmidrule(lr){18-20}\cmidrule(lr){21-21}
    & CR & RR & DR & AT & AT & & kbit/s & \multicolumn{2}{c}{ScoreQ} & ViSQOL$\uparrow$ & \multicolumn{2}{c}{ScoreQ} & ViSQOL & \multicolumn{2}{c}{ScoreQ} & ViSQOL$\uparrow$& \multicolumn{2}{c}{ScoreQ} & ViSQOL$\uparrow$& \textbf{RT60}\\
    \cmidrule(lr){9-10}\cmidrule(lr){11-11}\cmidrule(lr){12-13}\cmidrule(lr){14-14}\cmidrule(lr){15-16}\cmidrule(lr){17-17}\cmidrule(lr){18-19}\cmidrule(lr){20-20}
& & & & SS & DS & & & NR$\uparrow$ & REF$\downarrow$ & & NR$\uparrow$& REF$\downarrow$ & & NR$\uparrow$& REF$\downarrow$ & & NR$\uparrow$& REF$\downarrow$ & &\\
    \midrule
    Omran et al. \cite{omran2023disentangling} & & \checkmark & \checkmark & & \checkmark & 4 & 2.98 & - & - & - & \scorebar{2.74} & - & - & \scorebar{2.89} & - & - & \scorebar{2.44} & - & - & - \\
    \midrule
    Clean Recon & \checkmark & & & & & - & - & \textbf{\scorebar{4.37}} & \textbf{\refbar{0.10}} & \textbf{\scorebar{4.41}} & \scorebar{3.10} & \refbar{0.15} & \textbf{\scorebar{4.38}} & \scorebar{1.32} & \refbar{1.42} & \scorebar{1.07} & \scorebar{1.33} & \refbar{1.13} & \scorebar{1.08} & \corrbar{0.34}\\
    + Reverb Recon & \checkmark & \checkmark & & & & - & - & \scorebar{4.26} & \refbar{0.15} & \scorebar{4.30} & \textbf{\scorebar{3.12}} & \textbf{\refbar{0.14}} & \scorebar{4.32} & \scorebar{1.40} & \refbar{1.40} & \scorebar{1.13} & \scorebar{1.40} & \refbar{1.09} & \scorebar{1.12} & \corrbar{-0.01}\\
    + Dereverberation & \checkmark & \checkmark & \checkmark & & & - & - & \scorebar{4.30} & \refbar{0.13} & \scorebar{4.20} & \scorebar{3.05} & \refbar{0.17} & \scorebar{4.15} & \textbf{\scorebar{3.75}} & \textbf{\refbar{0.47}} & \scorebar{2.67} & \scorebar{2.99} & \refbar{0.74} & \scorebar{1.55} & \corrbar{0.88}\\
    \midrule
    AT Omran Taskset & & \checkmark & \checkmark & & \checkmark & - & - & \scorebar{4.12} & \refbar{0.22} & \scorebar{4.17} & \scorebar{3.01} & \refbar{0.17} & \scorebar{4.22} & \scorebar{3.62} & \refbar{0.52} & \textbf{\scorebar{2.71}} & \textbf{\scorebar{3.03}} & \refbar{0.40} & \scorebar{2.24} & \corrbar{-0.64}\\
    AT all tasks & \checkmark & \checkmark & \checkmark & \checkmark & \checkmark & - & - & \scorebar{4.14} & \refbar{0.21} & \scorebar{4.18} & \scorebar{2.99} & \refbar{0.21} & \scorebar{4.11} & \scorebar{2.69} & \refbar{0.94} & \scorebar{2.45} & \scorebar{2.91} & \refbar{0.47} & \scorebar{2.96} & \corrbar{-0.77} \\
    AT only AT & & & & \checkmark & \checkmark & - & - & \scorebar{3.96} & \refbar{0.29} & \scorebar{4.05} & \scorebar{2.99} & \refbar{0.20} & \scorebar{4.07} & \scorebar{2.99} & \refbar{0.86} & \scorebar{1.60} & \scorebar{2.95} & \textbf{\refbar{0.32}} & \textbf{\scorebar{3.02}} & \corrbar{0.89}\\
    \midrule
    AT Quantized & & \checkmark & \checkmark & & \checkmark & 4 & 4.0 & \scorebar{3.63} & \refbar{0.49} & \scorebar{2.65} & \scorebar{2.89} & \refbar{0.31} & \scorebar{2.49} & \scorebar{3.50} & \refbar{0.61} & \scorebar{2.05} & \scorebar{2.91} & \refbar{0.47} & \scorebar{1.64} & \corrbar{0.93}\\

    AT Quantized & & \checkmark & \checkmark & & \checkmark & 8 & 8.0 & \scorebar{3.82} & \refbar{0.38} & \scorebar{3.37} & \scorebar{2.95} & \refbar{0.25} & \scorebar{3.52} & \scorebar{3.59} & \refbar{0.54} & \scorebar{2.48} & \scorebar{2.99} & \refbar{0.43} & \scorebar{2.05} & \corrbar{-0.86}  \\
    AT Quantized & & \checkmark & \checkmark & & \checkmark & 16 & 16.0 & \scorebar{3.88} & \refbar{0.35} & \scorebar{3.47} & \scorebar{2.95} & \refbar{0.23} & \scorebar{3.66} & \scorebar{3.53} & \refbar{0.57} & \scorebar{2.47} & \scorebar{2.97} & \refbar{0.43} & \scorebar{2.03} & \corrbar{-0.68}\\
    \bottomrule
    \end{tabular}}
\end{table*}
\begin{figure*}[h]
\centering
    \begin{minipage}[t]{0.32\textwidth}
        \centering
        \includegraphics[width=\textwidth]{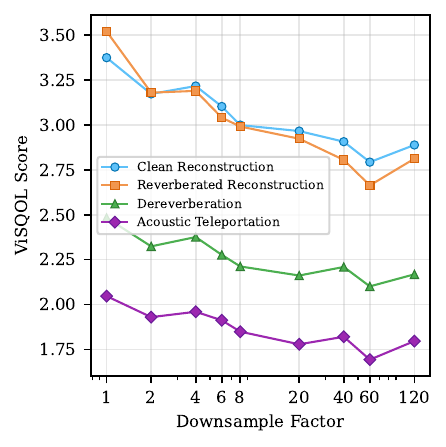}
        \captionof{figure}{\textit{Ablation on the effect of temporal downsampling of the acoustic embedding, $N=8$, Omran taskset.}}
        \label{fig:temporal_downsampling}
    \end{minipage}
    \hfill
    \begin{minipage}[t]{0.32\textwidth}
        \centering
        \includegraphics[width=\textwidth]{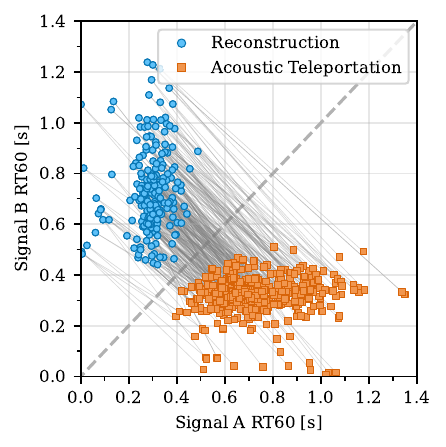}
        \captionof{figure}{Estimated RT60 for paired inputs and outputs after encoding and decoding with swapped acoustic embeddings, $N=8$, Omran taskset.}
        \label{fig:paired_inputs_outputs}
    \end{minipage}
    \hfill
    \begin{minipage}[t]{0.32\textwidth}
        \centering
        \includegraphics[width=\textwidth]{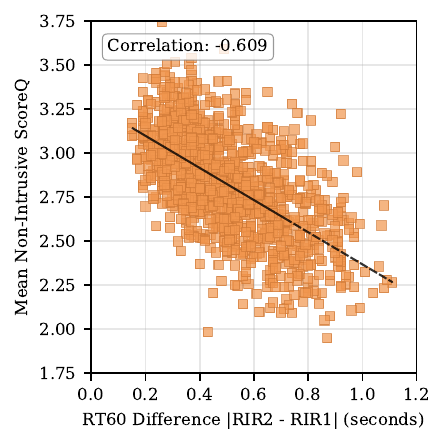}
        \captionof{figure}{\textit{Absolute RT60 difference before and after acoustic teleportation and measured mean ScoreQ NR value for both output signals, $N=8$, Omran taskset.}}
        \label{fig:rt60_quality}
    \end{minipage}
    \vspace{-1em}
\end{figure*}

\subsection{Datasets}
The speech data used for training and evaluation is sourced from DNS5 \textit{read\_speech} \cite{dubey2023icassp}, which is assumed to be anechoic. \acp{RIR} are sourced from GWAsmall \cite{tang2022gwa} containing simulated \acp{RIR}, exluding all \acp{RIR} with mean RT60 $>$ \SI{1.2}{s}. \acp{RIR} are preprocessed by removing pre-echoes, normalized by maximum absolute value, and scaled by 0.25 to prevent clipping during convolution. Following Omran et al. \cite{omran2023disentangling}, a balanced dataset is created, which helps to make the task of acoustic teleportation more explicit. This is done by selecting two \acp{RIR} per training sample: first with mean RT60 $<$ \SI{0.25}{s} and second with \SI{0.4}{s} $<$ mean RT60 $<$ \SI{1.2}{s}. For constructing the datasets this effectively limits $c_i$ and $r_j$ to $i,j \in \{1,2\}$. Reverberated speech is generated through convolution and normalized to the ±1 range. 

Data is organized into sample groups, each comprised of two three-second utterances of anechoic speech convolved with the two selected \acp{RIR}, yielding six signals per group, where reverberated output is trimmed to \SI{3}{s}. The complete dataset contains \SI{480000}{} training sample groups, providing \SI{400}{h} of clean speech and \SI{800}{h} of reverberated speech for training. The validation and test collections each contain \SI{1200}{} sample groups (\SI{2}{h} of clean and \SI{4}{h} of reverberated speech). Speakers and rooms are mutually exclusive across train/validation/test partitions.

Training follows FunCodec parameters \cite{du2023funcodec}, where weights for reconstruction loss and multi-spectral reconstruction loss are changed from 1.0 to 0.1 to re-balance the discriminator given the increased task complexity. All models were trained for 60 epochs on eight NVIDIA A100 GPUs.

\section{Evaluation}
\subsection{Objective Metrics}
Models are evaluated on the test set using ScoreQ \cite{ragano2024scoreq} and ViSQOL \cite{chinen2020ViSQOLv3opensource} metrics, with results reported in Table \ref{tab:performance_comparison}. ScoreQ provides both non-reference (NR) and reference-based quality assessment, while ViSQOL offers perceptual quality evaluation aligned with human auditory perception. To establish performance baselines for the proposed method, non-quantized task-specific networks are trained for clean reconstruction, reverb reconstruction, and dereverberation tasks. The clean reconstruction baseline achieves the highest performance with ScoreQ NR of 4.37 and ViSQOL of 4.41. The reverb reconstruction baseline maintains high quality (ScoreQ NR: 3.12, ViSQOL: 4.32), while the dereverberation baseline demonstrates the difficulty of this task with lower ViSQOL scores (ScoreQ NR: 3.75, ViSQOL: 2.67). All teleportation networks outperform the original Omran approach \cite{omran2023disentangling} in ScoreQ NR, with our AT Quantized models achieving scores of 2.91-3.03 compared to Omran's 2.44. 

The choice of training tasks impacts model performance across different applications. The AT-only configuration (AT-SS, AT-DS) excels in acoustic teleportation (ScoreQ NR: 2.95, ViSQOL: 3.02), but degrades clean reconstruction performance (ScoreQ NR: 3.96 vs. 4.14 all tasks training), indicating task-specific overfitting. Conversely, the Omran task set (RR, DR, AT-DS) achieves balanced performance, yielding competitive dereverberation results (ScoreQ NR: 3.62) while providing the best acoustic teleportation quality (ScoreQ NR: 3.03). The all-tasks configuration demonstrates moderate performance across all tasks but fails to excel in any specific application, suggesting potential interference between competing objectives.

Regarding quantization, analysis reveals diminishing returns beyond $N=8$ quantizers. Performance improvements from $N=4$ to $N=8$ are substantial (clean reconstruction: ScoreQ NR: 3.63 vs 3.82, acoustic teleportation: ScoreQ NR: 2.91 vs 2.99), while $N=8$ to $N=16$ yields minimal gains (ScoreQ NR: 3.82 vs 3.88). Despite saturation at $N=16$, a notable quality gap remains between quantized and non-quantized models (ScoreQ NR: 3.88 vs. 4.12), indicating suboptimal quantization for disentangled representations.

\subsection{Correlation with RT60}
To evaluate the interpretability of the learned acoustic embeddings, we aim to measure the correlation between those embeddings and RT60. To this end, we rely on 10-dimensional \ac{PCA} to project the embeddings to a lower-dimensional space, which is then used to calculate the correlation with the acoustic parameters. To account for overfitting, the PCA and standardization are fitted on tokens extracted from a training set, whereas the evaluation is conducted by applying the fitted PCA to tokens extracted from a test set. 

Pearson correlation coefficients are calculated between test set embeddings projected onto the first PCA component and RT60 averaged across frequency bands, derived from the normalized \acp{RIR} used for the original item generation. Results in Table~\ref{tab:performance_comparison} show strong correlations (often $>$ 0.6 and up to 0.93). The signs of the correlation are random across model configurations, which indicates variable embedding space orientation.

\subsection{Temporal Downsampling of Acoustic Embedding}

Investigating the effect of temporal downsampling and its viability as a mechanism to control the bitrate of acoustic embeddings, we evaluate models with different downsampling factors applied to the acoustic embedding while maintaining the speech embedding at full temporal resolution. We vary the downsampling factor from 1 (no downsampling) to 120 (downsampling to a single time frame). 

Figure~\ref{fig:temporal_downsampling} presents the ViSQOL scores for quantized models ($N=8$) across different downsampling factors for all individual evaluation tasks. The results show that higher downsampling factors lead to decreased ViSQOL scores across all tasks.  We conducted two-sample t-tests comparing each downsampling condition against the baseline (factor=1) to assess statistical significance, with sample sizes ranging from 756 to 3,019 per task. For all tasks, downsampling by even a factor of two leads to significantly worse performance ($p < 0.01$) compared to no downsampling.

\subsection{Acoustic Teleportation Evaluation}

Following the methodology in \cite{omran2023disentangling} to quantify the accuracy of acoustic teleportation in terms of reverberation time, we process input audio pairs $(x_{1,1}, x_{2,2})$ through a trained model ($N=8$) to obtain the reconstructed signals $(\hat{x}_{1,1}, \hat{x}_{2,2})$. RT60 values are estimated for all signals using a proprietary RT60 estimator, which demonstrates strong performance on the test set with RMSE = 0.058, Pearson correlation = 0.983, bias = 0.007, standard deviation = 0.057, and MAE = 0.032 when compared against ground truth RT60 values derived from the mean RT60 of the \acp{RIR} used to construct each signal. Then, we encode both waveforms and swap their acoustic embedding partitions to decode teleported signals $\hat{x}_{1,2}, \hat{x}_{2,1}$, and estimate their RT60 values. Figure~\ref{fig:paired_inputs_outputs} demonstrates successful RT60 swapping across most pairs, confirming that the acoustic embeddings capture the majority of RT60 information. Figure~\ref{fig:rt60_quality} shows that teleportation quality decreases as the RT60 difference between rooms increases, with a Pearson correlation of -0.61. This relationship holds across all model configurations.

\subsection{Disentanglement Performance Evaluation}

\begin{figure}[!t]
\centering
\begin{minipage}[!t]{0.24\textwidth}
    \centering
    \includegraphics[width=\textwidth]{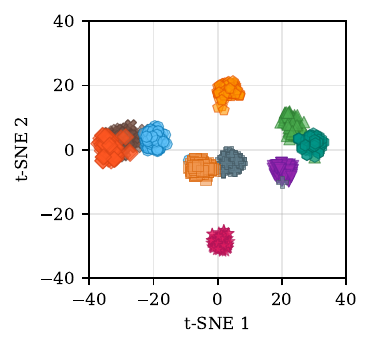}\\[-0.5em]
    \textbf{(a)} Room clusters from acoustic embedding
\end{minipage}%
\hfill%
\begin{minipage}[!t]{0.24\textwidth}
    \centering
    \includegraphics[width=\textwidth]{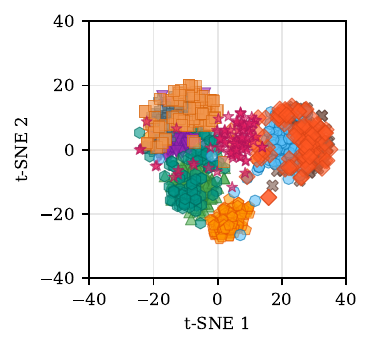}\\[-0.5em]
    \textbf{(b)} Room clusters from speech embedding
\end{minipage}%
\hfill%
\begin{minipage}[!t]{0.24\textwidth}
    \centering
    \includegraphics[width=\textwidth]{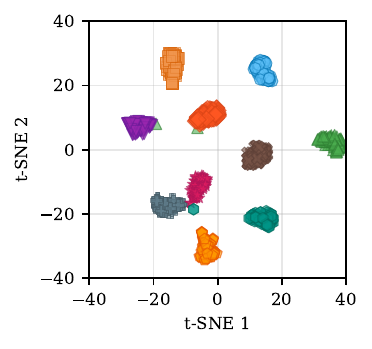}\\[-0.5em]
    \textbf{(c)} Speaker clusters from speech embedding
\end{minipage}%
\hfill%
\begin{minipage}[!t]{0.24\textwidth}
    \centering
    \includegraphics[width=\textwidth]{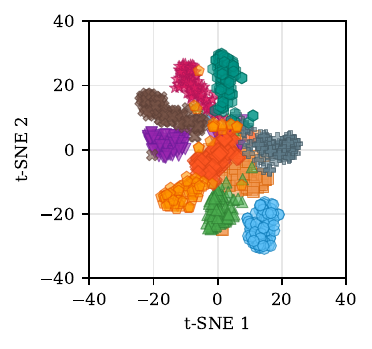}\\[-0.5em]
    \textbf{(d)} Speaker clusters from acoustic embedding
\end{minipage}
\caption{\textit{t-SNE clustering for 10 speakers in 100 rooms, clustered by speakers: from acoustic embeddings (a) and from speech embeddings (b). t-SNE clustering for 100 speakers in 10 rooms clustered by rooms: from speech embeddings (c) and from acoustic embeddings (d), $N=8$, Omran taskset.}} \vspace{-1em}
\label{fig:room_clustering}
\end{figure}

To quantify the disentanglement quality, we evaluate the dependency between acoustic and speech embeddings using t-SNE clustering analysis. For speaker independence, we select 10 diverse \acp{RIR} (RT60: \SI{0.04}{s} - \SI{2}{s}, C50: \SI{-9.8}{dB} - \SI{57.5}{dB}, DRR: \SI{-23.4}{dB} - \SI{15.6}{dB}) and convolve 100 random \SI{3}{s} DNS5 excerpts from the test set partition with each \ac{RIR}. These items are encoded by the quantized model ($N=8$), the acoustic and the speech embedding are extracted and temporally averaged. We then apply t-SNE visualization with perplexity=50, n\_iter=1000, random\_state=42. Figures~\ref{fig:room_clustering} (a) and (b) show the embedding clusters where each color corresponds to a given \ac{RIR}. Figure~\ref{fig:room_clustering} (a) shows distinct clustering of rooms from acoustic embeddings, while Figure~\ref{fig:room_clustering} (b) shows overlapping, more diffuse clusters from speech embeddings when grouped by room, indicating acoustic embeddings encode room-specific information mostly independent of speaker identity, while some information leakage remains. 

For room independence, we select 10 speech utterances from 10 different speakers and convolve with 100 random test \acp{RIR}. Figures~\ref{fig:room_clustering} (c) and (d) depict the clustering w.r.t. speaker identity. The t-SNE visualization reveals distinct speaker clustering from speech embeddings (Figure~\ref{fig:room_clustering} (c)) and less defined, overlapping clusters from the acoustic embeddings (Figure~\ref{fig:room_clustering} (d)). 

The separation in appropriate embedding spaces and mixing in inappropriate spaces demonstrates effective disentanglement: acoustic embeddings tend to be speaker-invariant but room-discriminative, while speech embeddings tend to be room-invariant but speaker-discriminative, confirming the separation of content and environmental characteristics.

\section{Conclusion}
In this work, we presented an approach for acoustic teleportation using disentangled neural audio codec representations. By adopting EnCodec and extending the training strategy, we achieved significant improvements over the baseline. Our ablation study shows that temporal downsampling of the acoustic embedding significantly degrades objective performance. The learned acoustic embeddings correlate strongly with RT60 and demonstrate successful disentanglement as shown by t-SNE clustering analysis, with acoustic embeddings clustering by room and speech embeddings by speaker.

However, several limitations constrain practical deployment. The present evaluation is restricted to English speech and simulated reverberation (RT60 $<$ \SI{1.2}{s}), limiting generalizability to real-world acoustic environments, especially with background noise. Quality degradation increases substantially when RT60 differences exceed \SI{0.8}{s}, indicating fundamental limits for extreme acoustic transformation. Future work should extend evaluation to real-world recordings, analyze speaker preservation, multilingual datasets, and non-speech audio content.

\vspace{-1em}
\section{Acknowledgements}
The authors gratefully acknowledge the scientific support and HPC resources provided by the Erlangen National High Performance Computing Center (NHR@FAU) of the Friedrich-Alexander-Universität Erlangen-Nürnberg (FAU). The hardware is funded by the German Research Foundation (DFG).


\normalsize
\bibliographystyle{IEEEbib}
\bibliography{strings,refs}

\end{document}